\documentclass[sigplan,9pt]{acmart}

\usepackage[ruled,vlined,linesnumbered]{algorithm2e} 
\usepackage{listings,xcolor}
\usepackage{graphicx}
\usepackage{subcaption}

\newcommand{\fabriccrdtspace}{\textsc{FabricCRDT }}
\newcommand{\fabriccrdt}{\textsc{FabricCRDT}}

\newcommand{\fabricspace}{\textsc{Fabric }}
\newcommand{\fabric}{\textsc{Fabric}}

\begin{document}

\title[\fabriccrdt: A CRDT Approach to Permissioned Blockchains]{\fabriccrdt: A Conflict-Free Replicated Datatypes Approach to Permissioned Blockchains}

\titlenote{This paper is a minor revision of the work published at the 20th ACM International Middleware Conference (Middleware '19). DOI: https://doi.org/10.1145/3361525.3361540. In the published version, Section 7, there is an error in drawing the charts concerning the performance of Hyperledger Fabric (Figure 3 – 7). In this revision, we corrected these figures. The findings and conclusions of the original paper remain unchanged. Please cite the original Middleware conference version of the paper.}

\author{Pezhman Nasirifard}
\affiliation{
  \institution{Technical University of Munich}
}
\email{p.nasirifard@tum.de}

\author{Ruben Mayer}
\affiliation{
  \institution{Technical University of Munich}
}
\email{ruben.mayer@tum.de}

\author{Hans-Arno Jacobsen}
\affiliation{
  \institution{Technical University of Munich}
}
\email{}

\renewcommand{\shortauthors}{P. Nasirifard, et al.}

\begin{abstract}
With the increased adaption of blockchain technologies, permissioned blockchains such as \textsc{Hyperledger Fabric} provide a robust ecosystem for developing production-grade decentralized applications. However, the additional latency between executing and committing transactions, due to \fabric's three-phase transaction lifecycle of Execute-Order-Validate (EOV), is a potential scalability bottleneck. The added latency increases the probability of concurrent updates on the same keys by different transactions, leading to transaction failures caused by \fabric's concurrency control mechanism. The transaction failures increase the application development complexity and decrease \fabric's throughput. Conflict-free Replicated Datatypes (CRDTs) provide a solution for merging and resolving conflicts in the presence of concurrent updates. In this work, we introduce \fabriccrdt, an approach for integrating CRDTs to \fabric. Our evaluations show that in general, \fabriccrdtspace offers higher throughput of successful transactions than \fabric, while successfully committing and merging all conflicting transactions without any failures.
\end{abstract}

%\begin{CCSXML}
%<ccs2012>
%<concept>
%<concept_id>10010520.10010521.10010537</concept_id>
%<concept_desc>Computer systems organization~Distributed architectures</concept_desc>
%<concept_significance>300</concept_significance>
%</concept>
%</ccs2012>
%\end{CCSXML}

%\ccsdesc[300]{Computer systems organization~Distributed architectures}


%\acmYear{2019}\copyrightyear{2019}
%\setcopyright{acmcopyright}
%\acmConference[Middleware '19]{Middleware '19: Middleware '19: 20th International Middleware Conference}{December 8--13, 2019}{Davis, CA, USA}
%\acmBooktitle{Middleware '19: Middleware '19: 20th International Middleware Conference, December 8--13, 2019, Davis, CA, USA}
%\acmPrice{15.00}
%\acmDOI{10.1145/3361525.3361540}
%\acmISBN{978-1-4503-7009-7/19/12}

\settopmatter{printacmref=false}
\setcopyright{none}
\renewcommand\footnotetextcopyrightpermission[1]{}
\pagestyle{plain}
\settopmatter{printfolios=true}

\maketitle

\section{Introduction} \label{introduction}

Since the introduction of Bitcoin~\cite{bitcoin}, new blockchains have
been developed that provide their users with novel ways of storing
and validating transactions and data in a trustless
environment~\cite{ethereum, zcash}. However, a large number of
existing blockchains fall significantly behind existing distributed
databases concerning scalability and
throughput~\cite{need_blockchain}. This limitation is touted as a
fundamental cost for providing security and trust in a decentralized
and trustless environment. Unfortunately, the significantly lower
transaction throughput of blockchains such as Bitcoin and
Ethereum~\cite{ethereum} has been a severe obstacle to the widespread
adoption of these technologies. Although there have been several
scalability approaches introduced~\cite{scale_bc}, the public and
permissionless nature of these blockchains make it difficult to find a
good solution~\cite{difficult}.

For use cases where the identity of users and nodes are known, when
developing decentralized enterprise applications, permissioned
blockchains constitute a viable alternative. One of the most prominent
permissioned blockchains is \textsc{Hyperledger Fabric} \cite{mainfabric}, which offers significantly higher throughput and transactional guarantees in comparison to Bitcoin and
Ethereum while allowing the deployment of Turing complete
applications~\cite{performance}. \fabricspace follows a three-phase
Execute-Order-Validate (EOV) lifecycle for transactions. To ensure the
consistency of the ledger, \fabricspace uses an optimistic concurrency
control mechanism that enables concurrent updates. This mechanism is
similar to the technique used by several database systems for
increasing scalability and throughput~\cite{databasify2, couchDB, hanna}. Although the concurrency control mechanism is necessary for
ensuring the consistency of the ledger, it constitutes a scalability
bottleneck for \fabric. The added latency between executing and
committing a transaction in \fabricspace is on the order of hundreds of
milliseconds to seconds, which is significantly higher than the
latency for committing transactions in conventional databases.

The considerable latency between the start (execution) and end (commit) of a transaction increases the probability of the concurrent arrival and execution of conflicting
transactions, which can result in the failure of a large portion of
transactions in the network~\cite{databasify}. Once a transaction
fails, the only option for clients is to create a new transaction and
resubmit, which adds to the complexity of \fabricspace application
development. Therefore, providing a solution that enables \fabricspace to
manage the conflicting transactions internally without rejecting the
transactions can significantly improve the throughput of \fabricspace and
simplify the application development process.

Conflict-free Replicated Datatypes (CRDTs)~\cite{first_crdt} address a similar concurrent update problem among node replicas. For certain update scenarios, CRDTs provide solutions for merging conflicting values internally and resolving update conflicts automatically. CRDTs are abstract datatypes that allow node replicas to eventually converge to a consistent state without losing updates. CRDTs have been implemented in several production-grade databases such as Redis~\cite{redis} and Riak~\cite{Riak} and have established themselves as a viable solution in practice~\cite{second_crdt}. In this work, we introduce \fabriccrdt, an extension of \fabricspace with CRDTs, and we assess the improvements CRDTs introduce to permissioned blockchains.

In doing so, we offer the following contributions in this work:
\begin{enumerate}

\item We investigate the applicability of CRDTs to permissioned
  blockchains and propose a novel approach for integrating CRDTs with
  \fabricspace that automatically resolves transaction conflicts without the
  loss of updates.

\item We extend \fabricspace with CRDTs and describe the design of a new
  system called \fabriccrdt. Our extension is implemented without
  disrupting the standard behavior of \fabric, making it backward
  compatible with existing chaincodes and transactions. Also, our
  approach requires only  minimal changes to \fabricspace and reuses its main
  components.

\item Our solution simplifies the complexity of \fabricspace application
  development by eliminating the developer's concerns about
  transaction failures due to concurrent update conflicts. Also, we
  offer a simplified CRDT-based programming model for developing
  CRDT-enabled applications on \fabriccrdt, with a minimal learning
  curve for developers who are already familiar with \fabric.

\item We provide insights into the appropriate use cases for a
  CRDT-enabled permissioned blockchain. We also perform extensive
  evaluations to understand the best configuration of \fabriccrdtspace for
  executing CRDT-compatible applications.
\end{enumerate}

The remainder of the paper is organized as follows. First, we provide
an overview of \fabricspace and CRDTs in Section~\ref{background}, followed
by a detailed discussion of multi-version concurrency control (MVCC)
conflicts in Section~\ref{analysis}, and our approach in
Section~\ref{design_des} and Section~\ref{fabric_extensions}. We
discuss use cases and the potential of CRDT-enabled blockchains in
Section~\ref{crdts_on_fabric}. We report the results of our evaluation
in Section~\ref{evaluation} and review related work in
Section~\ref{related_work}.

\begin{figure*}
  \centering
  \includegraphics[width=.95\linewidth]{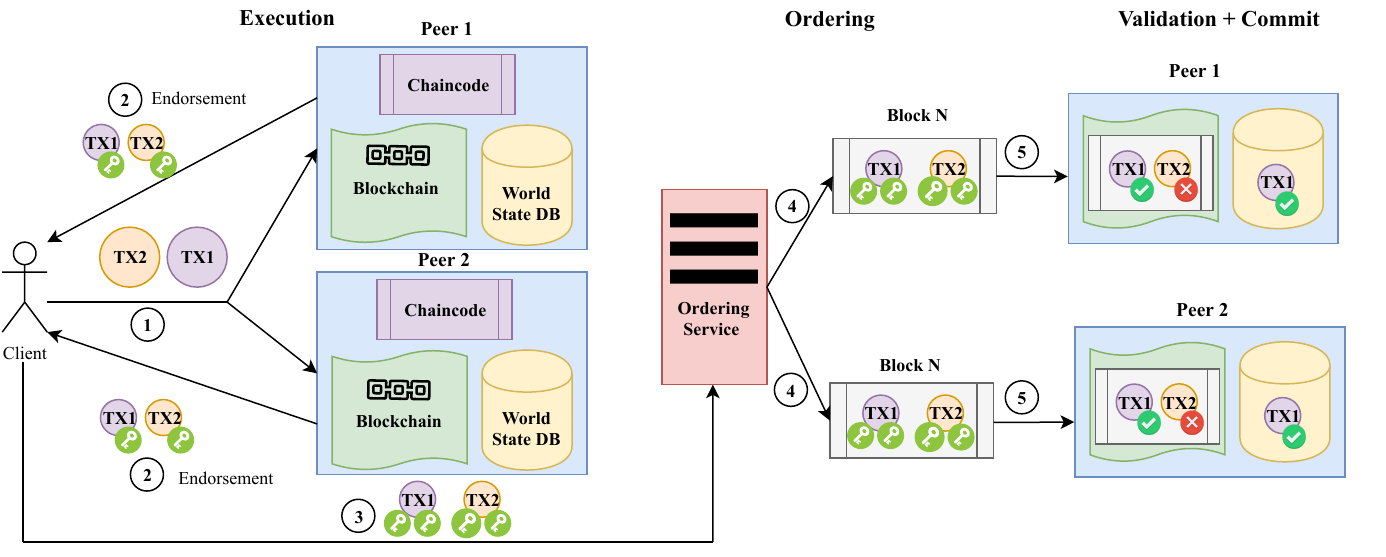}
  \caption{Transaction flow in \textsc{Hyperledger Fabric}.}
  \label{fig:fabtic_flow}
\end{figure*}

\section{Background} \label{background}

\subsection{\textsc{Hyperledger Fabric}} \label{fabric}

\textsc{Hyperledger Fabric} (\fabric) is an open-source permissioned blockchain,
initially developed by IBM~\cite{fabric_main_page}. \fabricspace provides an
ecosystem for hosting and executing blockchain applications, offering
a wide range of features and services ranging from storing the
application data to a sophisticated identity and membership
management. Developers host smart contracts, also known as chaincodes,
on \fabricspace which clients interact with by creating and submitting
transactions. Developers use \textit{chaincode shim} that provides
APIs for the chaincode to interact with data on the ledger. Shim is
available in programming languages such as Go and Node.js.

The two main components of \fabricspace are \textit{peers} and
\textit{orderers}. Peers are responsible for executing transactions
and storing the data on their local copy of the ledger. The peer's
ledger consists of an append-only blockchain and a world state
database, realized by CouchDB~\cite{couchDB}. Executing all valid
transactions included in the blockchain starting from the genesis
block results in the current state of the world state
database. Orderers are responsible for defining a total global order
for transactions and batching the ordered transactions into
blocks. \fabricspace divides peers into \textit{organizations} and provides
them with private communication channels. A complete workflow for
committing a transaction is depicted in Figure~\ref{fig:fabtic_flow}
(adapted from Ref.~\cite{databasify} ). Every successfully committed
transaction follows these three steps:

\begin{enumerate}

\item \textbf{Execution and Endorsement:} The client creates a
  \textit{transaction proposal} containing the name of the chaincode,
  the input data, and the endorsement policy. An endorsement policy
  specifies which peers from which organizations are required to
  execute and sign the proposal (also known as endorsing). The client
  submits the proposal to the peers specified by the endorsement
  policy in parallel (Step~1 in Figure~\ref{fig:fabtic_flow}). Each
  endorsing peer executes the chaincode against the local copy of the
  world state database, signs the results, and sends the results back
  to the client (Step~2). The results are in the form of read and
  write sets, where read sets contain the keys read during execution
  and write sets contain the key-value pairs to be written to the
  ledger. Peers do not modify their local copy of the ledger during
  this phase and only execute the proposal in an isolated fashion,
  i.e., peers \textit{simulate} the transaction proposal.

\item \textbf{Ordering:} Once the client has received enough
  endorsements that satisfy the endorsement policy, it creates a
  transaction containing the proposal's payload, the endorsements, and
  other metadata, and sends the transaction to the ordering service
  (Step~3). The ordering service receives transactions from every
  client in the network, defines a total global order for the
  transactions and batches them into new blocks (Step~4), which are
  broadcast to the peers (Step~5).

\item \textbf{Validation and Commit:} Peers perform two validations on
  the transactions in the incoming blocks and then commit the
  transactions. For the first validation, a peer validates the
  endorsement policies of the transactions in parallel to ensure that
  each transaction satisfies the predefined endorsement
  policy. Second, a peer sequentially compares the read-set with its
  local copy of the world state database to ensure that the records
  that were read during the endorsement phase have not changed
  concurrently in the world state database. The transactions that
  successfully pass both validations are considered valid. Finally, a
  peer appends every valid and invalid transaction in the block to the
  blockchain and updates the world state database with the write-set
  of the valid transactions.

\end{enumerate}

\subsection{Conflict-free Replicated Datatypes} \label{crdts}

CRDTs are abstract datatypes that can be replicated on several nodes
with the guarantee to eventually converge to the same state without
requiring a consensus protocol~\cite{first_crdt, second_crdt}. CRDTs
provide well-defined application interfaces, representing specific
data structures such as counters, sets, lists, maps, JSON objects, and
others. CRDTs provide these interfaces by extending the basic
datatypes with some metadata, which makes the updates on these datatypes at least commutative. Thus, commutative updates can be applied
in different orders on replicas, resulting in the same state no matter
the order of updates, provided no updates are lost or duplicated.

In general, CRDTs are divided into two main categories: state-based
CRDTs and operation-based CRDTs. State-based CRDTs exchange the full
state or delta state and merge the local state with the received
state. For operation-based CRDTs, nodes propagate the state by sending
update operations to other nodes. As an example, a counter datatype
which only increments a value by one can be converted to a grow-only CRDT
counter, by defining an \textit{increment operation} that increments
the value of the counter by one. The grow-only CRDT counter is
relatively easy to create since the increment operation is inherently
commutative, although not idempotent. Therefore, the grow-only CRDT
counter converges to the same state no matter in which order the
operations are applied, yet the same operation cannot be applied more
than once. For interacting with this counter over the network, nodes
send increment operations, and given an asynchronous distributed
system where the delivery of messages eventually succeeds without
message loss nor duplication, the counter eventually converges to the
same value on all nodes.

Among existing CRDTs, a JSON CRDT represents a complex general-purpose
data structure~\cite{jsoncrdt}. A JSON object is inherently a tree
structure, consisting of other structures like maps and lists. In
JSON, a map is a dictionary of key-value pairs where keys are string
constants and values are either primitive values like string or
numbers or complex structures like other maps and lists. In this work,
we assume that maps are unordered structures and the values in the
maps are either a string, a map, or a list. In JSON, lists are arrays
of objects, which can be a combination of primitive values or complex
structures, like string, numbers, maps or lists.

\section{MVCC Conflicts} \label{analysis}

In this section, we analyze \fabric's concurrency control mechanism and
the causes of transaction failures in more detail.

A transaction proposal invokes the chaincode on \fabric, which, based
on the implemented logic, interacts with the stored ledger data in
three ways during the chaincode execution:

\begin{itemize}

\item \textbf{Read Transaction:} Chaincode only reads key-value pairs
  from the ledger.

\item \textbf{Write Transaction:} Chaincode only writes key-value
  pairs to the ledger without reading any pair.

\item \textbf{Read-Write Transaction:} Chaincode reads and writes
  key-value pairs from/to the ledger.

\end{itemize}

The execution of transaction proposals results in a read-write
set. The read set includes a list of keys and the version number of
the key's value that a peer retrieved from the ledger during the
execution of the chaincode. The write set contains the key-value pairs
that will be committed to the ledger at the end. Read transactions do
not change the state of the ledger, and clients do not send the
transactions for ordering and committing. The write transaction
results in a read-write set with an empty read set. Hence, these
transactions will not cause any read-write set conflict. However, a
read-write transaction with an outdated version number in the read-set
fails the validation. To illustrate the problem better, imagine that
at time \textit{TS\textsubscript{1}}, peer \textit{P\textsubscript{1}}
has the world state \textit{WS: \{ $(K_{1}, \textit{VN}_{1}, \textit{VL}_{1})$, $(K_{2},
  \textit{VN}_{2}, \textit{VL}_{2})$, $(K_{3}, \textit{VN}_{3}, \textit{VL}_{3})$ \}} and receives a block
containing five transactions with corresponding read-write sets as
follows, where \textit{K} represent the key of the key-value pairs,
\textit{VN} represents the version number of the retrieved key-value
pairs from the world state and \textit{VL} is the value of the key to
be written to the blockchain and world state:

\begin{itemize}

\item \textbf{T\textsubscript{1}:} $< \textbf{Read}: \{(K_{2}, \textit{VN}_{2})\}, \textbf{Write}: \{(K_{2}, \textit{VL}_{1})\} >$

\item \textbf{T\textsubscript{2}:} $< \textbf{Read}: \{(K_{1}, \textit{VN}_{1}), (K_{2}, \textit{VN}_{2})\}, \textbf{Write}: \{(K_{3}, \textit{VL}_{3})\} >$

\item \textbf{T\textsubscript{3}:} $< \textbf{Read}: \{(K_{2}, \textit{VN}_{2})\}, \textbf{Write}: \{(K_{3}, \textit{VL}_{1})\} >$

\item \textbf{T\textsubscript{4}:} $< \textbf{Read}: \{(K_{3}, \textit{VN}_{2})\}, \textbf{Write}: \{(K_{2}, \textit{VL}_{1})\} >$

\item \textbf{T\textsubscript{5}:} $< \textbf{Read}: \{\}, \textbf{Write}: \{(K_{3}, \textit{VL}_{2})\} >$

\end{itemize}

Given that all five transactions pass the endorsement policy
validation (not explicitly shown here), \textit{P\textsubscript{1}}
sequentially validates the five transactions in the block by comparing
the version number of each key in the read set to the version number
in the world state. A transaction is considered valid if both version
numbers are equal. If the version numbers are not equal, the peer invalidates
the transaction as a \textit{multi-version concurrency control
  conflict (MVCC conflict)}. The key's mismatch is the result of
updates committed by preceding valid transactions. The preceding
transactions may be included either in the previous blocks or in the
same block but are preceding the current position of the conflicting
transaction. Committing keys in the write-set of the valid
transactions causes the version number of keys in the world state
database to change. Therefore, \textit{P\textsubscript{1}} marks
T\textsubscript{1} as valid and T\textsubscript{2} and
T\textsubscript{3} as invalid, because the write-set of
T\textsubscript{1} updates K\textsubscript{2} so that its new version number $\textit{VN}_{2}'$
and the version number of K\textsubscript{2} in T\textsubscript{2} and
T\textsubscript{3}'s read-set does not match ($\textit{VN}_{2} \neq  \textit{VN}_{2}'$). Also, P\textsubscript{1}
marks T\textsubscript{4} and T\textsubscript{5} as valid.

This multi-version concurrency control mechanism is a commonly used
method in database systems to increase the throughput and decrease the
latency, instead of using blocking mechanisms such as shared
locks~\cite{databasify2}. Although this mechanism is necessary for
ensuring data consistency and isolation of transactions, the relatively high latency between the
creation of the read-write set and the validation of the read-set in
\fabricspace can result in a large number of transactions in a block to
fail, especially when a small set of frequently accessed keys are
included~\cite{databasify, xox}. This high latency consists of the
endorsement latency, the ordering latency, and the commit
latency~\cite{performance}. The endorsement latency is the time needed
for the client to obtain all the required endorsements, which,
depending on the endorsement policy and the complexity of chaincodes,
varies significantly for different transactions. The ordering latency
is the time required for the transaction to be included in one block
and to be broadcasted to the peers. The ordering service creates a
block based on several criteria, including the maximum number of
transactions, the maximum total size of transactions in a block and
a timeout period for creating blocks. For higher transaction arrival
rates, the ordering service creates a block as soon as the maximum
size is reached. However, for lower arrival rates, the transaction can
be delayed until the timeout period is reached. The timeout period is a
configurable parameter which can be on the order of seconds. Finally, the
commit latency is the time taken for a peer to validate and commit
transactions in the block. These delays are inherent to the design of
\fabricspace and can not be significantly reduced without fundamental
changes to \fabric.

\section{\fabriccrdtspace Design} \label{design_des}
 
In this section, we discuss the system model and requirements of \fabriccrdtspace and introduce our approach for integrating CRDTs with \fabric.
 
\subsection{System Model} \label{fabric_system_model}
 
 As \fabriccrdtspace is an extension of \fabric, we assume the same system model as \fabric. We consider an asynchronous distributed model where all users and nodes are connected in a way that it is guaranteed for all transactions to be delivered eventually, despite arbitrary delays. However, the order of the transactions in a block is not guaranteed to be the same order that transactions are issued or arrive at the ordering service. As the ordering service does not guarantee to prevent duplicate delivery of transactions, for example, when the client intentionally or unintentionally submits duplicate transactions, we rely on peers to identify the duplicate ones with same transaction identification numbers. In this work, we assume that clients do not submit duplicate transactions. In the case that duplicate transactions are submitted, \fabriccrdtspace also commits duplicate transactions.

\subsection{\fabriccrdtspace Requirements} \label{fabric_requirments}

We define a set of requirements that \fabriccrdtspace must satisfy. First, \textit{compatibility}: we aim to extend \fabricspace with CRDT-enabled functionalities with minimal changes to the original design of \fabric. This way, we keep the learning curve minimal for developers who already designed applications for \fabric. Also, the applications developed for \fabricspace remain compatible with \fabriccrdt. The second requirement is \textit{no failure}: \fabriccrdtspace should be able to commit all valid CRDT-based transactions successfully. We define valid transactions as the transactions submitted by the client which pass the endorsement policy validation successfully. The third requirement is \textit{no update loss}: by committing all valid transactions in a block, \fabriccrdtspace eventually converges to the same state on all peers, and all client's updates are preserved while using CRDT techniques to merge conflicting transactions. The last requirement is \textit{generality}: to accommodate developers with the possibility of realizing a wide range of use cases, the CRDT approach used for merging conflicting transactions in \fabriccrdtspace should provide users with a general-purpose data structure to submit data to the ledger.

\begin{figure}
  \centering
  \includegraphics[width=1\linewidth]{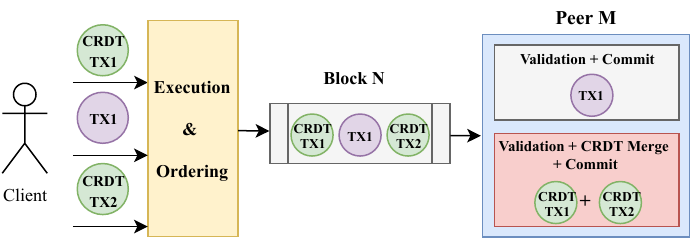}
  \caption{Transaction flows on \fabriccrdt.}
  \label{fig:fabticcrdt_flow}
\end{figure}

\subsection{\fabricspace and CRDTs} \label{crdts_on_fabric_imp}

To achieve the requirement we discussed, we need to identify the right approach for dealing with conflicting transactions internally. As discussed in Section~\ref{analysis}, \fabricspace rejects transactions with an outdated version number of key-value pairs in the read-set and discards these transactions' write-set, as committing a write-set with outdated version can result in data inconsistencies. To avoid the failure of conflicting transactions and data inconsistencies, \fabriccrdtspace does not reject transactions, but merges the values of the conflicting transactions by using CRDT techniques.

Since we aim to keep \fabricspace applications compatible with \fabriccrdt, we define a new type of transaction that encapsulates all CRDT-related functionalities. Figure~\ref{fig:fabticcrdt_flow} displays the transaction flow in \fabriccrdt, where CRDT and non-CRDT transactions coexist. CRDT transactions have a similar structure as standard \fabricspace transactions; however, they invoke chaincodes which modify CRDT-encapsulated values on the ledger. The chaincodes that contain CRDTs are executed in the same way as non-CRDT chaincodes, but peers flag the key-value pairs in the resulting transaction's write-set as ``CRDT key-values''. On \fabriccrdt, both types of transactions go through the same ordering steps, but they are treated differently in the final validation and commit phase. Non-CRDT transactions go through the same validation steps as on \fabric, but CRDT transactions only go through the endorsement validation check. Then, instead of going through the MVCC validation, the transaction values of conflicting transactions are merged automatically by using CRDT techniques before being committed to the ledger. The CRDT procedures used for conflict resolution depends on the type of CRDT object; for example, managing grow-only CRDT counters requires different techniques than merging JSON CRDTs. In our prototype of \fabriccrdt, we support merging JSON CRDTs. 

\section{\fabriccrdtspace Implementation} \label{fabric_extensions}

In the following section, we discuss the implementation of \fabriccrdtspace in detail. We introduce our approach to integrate CRDTs with peers. We also explain the mechanism for merging JSON objects using CRDT techniques. We implemented \fabriccrdtspace based on \fabricspace v1.4.0.

\begin{algorithm}[h]
\SetKwInOut{Input}{input}\SetKwInOut{Output}{output}
 \caption{Merge CRDT transactions in a block.}
   \label{alg:crdtalg}
 \SetKwProg{ValidateMergeBlock}{ValidateMergeBlock}{}{}
 \ValidateMergeBlock{$(\textit{Block})$}{
    \Input{$\textit{Block}$, A block received from the orderer.}
    \Output{$\textit{MergedCRDTsBlock}$, A block with merged CRDT transactions to be committed to the ledger.}
    $\textit{CRDTs} = \textit{set}()$ 
    
    \ForEach{$\textit{TX}_i \in \textit{Block.Transactions}$}{
    
     	\ForEach{$\textit{key}_j,  \textit{value}_j \in \textit{TX}_i.\textit{WriteSet}$}{
     	
				\If{$\textit{value}_j.\textit{IsCRDTObject}()$}{
				
      				$\textit{value}_j.\textit{SkipMVCCValidation}()$
      				
      				$\textit{CRDT} = \textit{CRDTs.GetObjectIfExists}(\textit{key}_j)$
      				
      					\If{$\textit{CRDT} == \textit{Null}$}{
      					
      							$\textit{CRDT} = \textit{InitEmptyCRDT}(\textit{key}_j, \textit{value}_j)$
      							
      							$\textit{CRDTs.SetObject(CRDT)}$
      							
      					}
      				
      				$\textit{MergeCRDT(CRDT, value}_j)$
      				
      				$\textit{CRDTs.SetObject(CRDT)}$
      
      			}     	
      			\Else{
      					// Skip the key-value and let it be handled as non-CRDT transactions.
  				}
  		  }
    }
    
    $\textit{DoMVCCValidationOnNonCRDTTransactions(Block)}$
    
     \ForEach{$\textit{TX}_i \in \textit{Block.Transactions}$}{
     
     	\ForEach{$\textit{key}_j,  \textit{value}_j \in \textit{TX}_i.\textit{WriteSet}$}{
     	
				\If{$\textit{value}_j.\textit{IsCRDTObject()}$}{
				      				
      				$\textit{CRDT} = \textit{CRDTs.GetObjectIfExists(key}_j)$
      				
					$\textit{DataTypeObject} = \textit{CRDT.ConvertCRDTToDataType()}$  
					    				      				
      				$\textit{value}_j = \textit{DataTypeObject.ConvertToBinary()} $
      				
      				$\textit{TX}_i.\textit{UpdateWriteSet(key}_j, \textit{value}_j)$ 
      				
      			}     	
  		  }
    }
    
    \KwRet{$\textit{Block}$}
  }
 \end{algorithm} 
 
\subsection{CRDT Transactions in a Block} \label{crdt_transaction_merge_valid}

The CRDT and non-CRDT transactions in a block follow the same workflow until they reach the multi-version concurrency control validation (MVCC validation). The non-CRDT transactions pass through the MVCC validation, and the peer commits the valid transactions to the ledger. The CRDT transactions in a block are merged before getting committed. Algorithm~\ref{alg:crdtalg} explains our approach for managing transactions in a block on \fabriccrdt.

For resolving the CRDT transactions in one block, first, we iterate through all transactions in the block, and for each transaction, we iterate through the key-value pairs in the transaction's write-set (lines 3 to 14 in the algorithm). If the key-value pair is not marked as a CRDT, we skip the key-value pair to be handled as a non-CRDT transaction. However, if the key-value pair is flagged as a CRDT, the algorithm first checks if a CRDT object with the same key already exists in a local set containing all CRDT objects. If a CRDT object does not exist, the algorithm instantiates a new CRDT object with the key and adds it to the set. The type of CRDT object depends on the type of CRDT value in the key-value pair. For example, for a JSON CRDT type, an empty JSON CRDT object is instantiated. Afterward, the peer converts the binary value of the key-value pair to the corresponding type and merges it with the CRDT object. Then, the set containing all CRDT objects is updated (lines 7 to 12). We discuss the steps required for merging the individual CRDTs in Section~\ref{crdt_json_struct}. After the first iteration, the peer performs MVCC validation on non-CRDT transactions (line 15). Afterward, the algorithm iterates through every transaction's write-set once more to check if there exists a CRDT object for that key in the local CRDT set (lines 16 to 22). If a CRDT object exists, then the CRDT object is converted to the corresponding datatype, for example, for a JSON CRDT, it is converted to a JSON object (line 20). The converted object is a representation of the datatype with all the CRDT-related metadata cleaned up and removed. Finally, the object is converted into a byte array that replaces the value of the key-value pair in the write-set of the transaction (lines 21 and 22). This second iteration through every transaction's write set is necessary because the peer is not aware of all key-value pairs in the CRDT transactions in the block that needs to be merged until the end of the first iteration. Once all CRDT transactions are merged, the peer finalizes and cleans up the CRDT objects and updates the write-values of the corresponding transactions with the new converged value, which is then committed to the ledger.

As an example, consider two JSON objects in the write-sets of two different transactions that have the same key, as depicted in Listing~\ref{lst:json_sample}. Since the values have JSON types, Algorithm~\ref{alg:crdtalg} creates one JSON CRDT with the identifier \textit{Device1} and extends and merges the created JSON CRDT with both values.

\begin{lstlisting}[basicstyle=\small,label=lst:json_sample,caption=Sample JSON objects in transactions' write-set.]
"CRDT-Transaction1-Write-Set" : [(
	"Key" : "Device1",
	"Value": {
	     "tempReadings": [{
	     	"temperature": "15"
}]})]
"CRDT-Transaction2-Write-Set" : [(
  	"Key" : Device1",
	"Value": {
	     "tempReadings": [{
	     	"temperature": "20"
}]})]
\end{lstlisting}

The result of merging the two CRDT transactions is shown in Listing~\ref{lst:json_sample_result}. The write-set of Transaction 2 is identical to the write-set of Transaction 1. 

\begin{lstlisting}[basicstyle=\small,label=lst:json_sample_result,caption=Result of example JSON merge.]
"CRDT-Transaction1-Write-Set" : [(
	"Key" : "Device1",
	"Value": {
         "tempReadings": [{
               "temperature": "15"
         }, {
               "temperature": "20"
}]})]
"CRDT-Transaction2-Write-Set" : [(
	"Key" : "Device1",
	"Value": {
         "tempReadings": [{
               "temperature": "15"
         }, {
               "temperature": "20"
}]})]
\end{lstlisting}

\subsection{JSON CRDTs on \fabriccrdt} \label{crdt_json_struct}

Although the approach discussed in Algorithm~\ref{alg:crdtalg} is independent of the CRDT types, the necessary mechanism for resolving conflicts and merging different CRDTs varies between different CRDT types and requires specific implementation support. In our prototype of \fabriccrdt, we focused on implementing and integrating JSON CRDTs~\cite{jsoncrdt}, which provide a general-purpose data structure for complex use cases.

We implemented JSON CRDTs based on the theoretical work of Klepmann et al.~\cite{jsoncrdt} and a GoLang JSON CRDT implementation~\cite{json_imp}. In Ref.~\cite{jsoncrdt}, the authors introduce the formal semantics and the algorithm for implementing an API for interacting with a JSON CRDT. The algorithm provides an API for modifying JSON objects, such as inserting, assigning, and deleting values, as well as reading from the JSON. The Reading API does not cause any modification to the JSON, but modifying the JSON is represented by \textit{operations} which have globally unique identifiers. The API described by the authors, although necessary for ensuring the automatic resolution among several processes, is cumbersome to use for chaincode developers. In \fabriccrdt, every peer observes the transactions in a block in the same order. We exploit this property to simplify the API. To use the JSON CRDTs in the chaincode, similar to chaincodes on \fabric, developers should create JSON objects. However, for submitting the key-value pairs to the ledger, the developer should use the CRDT-specific \textit{putCRDT} command that we implemented in the chaincode shim. This command only informs the peer that this value is a CRDT and does not interact with the CRDT in any way. The operations required for merging the JSON CRDTs are performed on the peers without the interference of the chaincode developer.
 
Algorithm~\ref{alg:crdtmeger} describes our approach for merging JSON CRDTs. This algorithm is the implementation of the \textit{MergeCRDT} function in line 11 of Algorithm~\ref{alg:crdtalg}. Algorithm~\ref{alg:crdtmeger} iterates through each key-value pair in the JSON object, where the value is either a string, a list, or a map. The items included in the list or map may include nested maps or lists. For each value in the JSON object, first, we create an empty cursor and an empty dependency list (lines 3 and 4). The cursor defines the path from the head of the JSON CRDT to the node where the mutation for modifying the JSON CRDT happens. A mutation defines the modification, such as add or delete, that is applied to the JSON object. The dependencies set contains the unique identifier of all operations which should be performed before the current operation is executed. We ensure that the operations identifiers are globally unique by using an instance of a Lamport Clock~\cite{lamport} for each JSON CRDT instantiation. The Lamport clock is incremented by one with every new operation to ensure the causal order of the operations. 

\begin{algorithm}[h]
\SetKwInOut{Input}{input}\SetKwInOut{Output}{output}
 \caption{Merge a JSON object with JSON CRDT.}
   \label{alg:crdtmeger}
 \SetKwProg{MergeCRDT}{MergeCRDT}{}{}
 \MergeCRDT{$(\textit{JsonCRDT, Json})$}{
    \Input{$\textit{JsonCRDT}$, An initialized JSON CRDT object.}
    \Input{$\textit{Json}$, A JSON object to be added to the JSON CRDT.}
   
   \ForEach{$\textit{key}_i,  \textit{value}_i \in \textit{Json}$}{
			
			$\textit{cursor} := \textit{NewCursorElements()}$   
			
			$\textit{dependencies} = \textit{set()}$
   
  			 \If{$\textit{value}_i.\textit{IsString()}$}{
  			 
  			 	$\textit{AddCursorElement(cursor, key}_i)$
  			 	
  			 	$\textit{JsonCRDT.TickClock()}$
  			 	
  			 	$\textit{mutation} = \textit{NewInsertMutation(key}_i, \textit{value}_i)$
  			 	
  			 	$\textit{operation} = \textit{NewOperation(JsonCRDT.ClockToString()}, $
  			 	$\textit{dependencies, cursor,  mutation)} $
  			 	
  			 	$\textit{ApplyOperationToJSON(JsonCRDT, operation)}$
  			 	
  			 	$\textit{dependencies.Add(JsonCRDT.ClockToString())}$
  			 
  			 } \ElseIf{$\textit{value}_i.\textit{IsList()}$}{
  			 
				\ForEach{$\textit{listValue}_j \in \textit{value}_i.\textit{GetListItems()}$}{
				
					$\textit{AddCursorElement(cursor, key}_i)$
					
					$\textit{RecursivelyAddListItemToJsonCRDT(}$
					$\textit{JsonCRDT, key}_i\textit{, listValue}_j, $
					$\textit{dependencies, cursor)}$
					
					$\textit{RemoveCursorElement(cursor, key}_i)$
					
				}  			 
  			 
  			 } \ElseIf{$\textit{value}_i\textit{.IsMap()}$}{
  			 
  			 \ForEach{$\textit{mapKey}_j\textit{, mapValue}_j \in \textit{value}_i\textit{.GetMapItems()}$}{
  			 
  			 		$\textit{AddCursorElement(cursor, key}_i)$
					
					$\textit{RecursivelyAddMapItemToJsonCRDT(}$
					$\textit{JsonCRDT, mapKey}_j\textit{, mapValue}_j,$
					$\textit{dependencies, cursor)}$
					
					$\textit{RemoveCursorElement(cursor, key}_i)$
				
				}  	
 
  			 }
       }
  }
 \end{algorithm} 

If the value of a key in the JSON object is a string, the algorithm executes lines 6 to 11. First, it extends the cursor with the current key and increments the Lamport clock by one. Then, it creates a mutation for inserting the current string value and the current key. Afterward, an operation is created with the current value of the Lamport clock as the identifier of the operation. The operation also holds the mutation, a dependency list, and the cursor pointing to the location in the JSON CRDT where the mutation occurs. Finally, the operation is applied to the JSON CRDT (line 10). For applying the operation, first, we check if all dependencies in the operation's dependency list are already applied. If some of the operations are missing, we queue the operation until all dependencies are applied. If there is no pending operation, we apply the operation by using the operation's cursor to traverse from the head of the JSON CRDT. For every node in the cursor, if the node already exists, we add the identifier of the current operation to the node to record the current operation's node dependencies. If the node from the cursor is missing in the JSON CRDT, we add the node to the JSON CRDT and the operation's identifier to the node. Once we reach the end of the cursor and the location of the node is found, we apply the mutation to the JSON CRDT and insert the node. For adding the node, we insert a dictionary item with the key as the operation identifier, and the value as the string value from the JSON object.

When the value of the JSON object is a list, we iterate through the list's items (lines 14 to 16). For every list item, first, we append the cursor with the current key in the JSON object, then we call a recursive function that extends the JSON CRDT with the content of the list item. We use a recursive function since the value of the list item could either be a string, a list, or a map, which may contain further nested list or map items. The recursive function either extends the JSON CRDT with the string value as described in the algorithm (lines 6 to 11) or if the value is a list or a map, it extends the JSON CRDT (lines 13 to 16 or lines 18 to 21, respectively.) When the value of the JSON object is a map (lines 18 to 21), we follow the same approach as the list type, but we extend the cursor with the key of the key-value pairs in the map instead of the key of the current JSON object. 

To limit the complexity of our prototype, the JSON lists in our system only support string, map, and list. Therefore, when users require to use other datatypes, such as numbers or Boolean, they should convert the desired datatype to strings.

\section{Use Cases and Potentials of a CRDT-enabled \fabric} \label{crdts_on_fabric}

Numerous use cases have been proposed that benefit from CRDT-enabled database systems, such as data metering, global voting platforms, and shared document editing applications~\cite{when_crdt, second_crdt}. Similarly, these use cases also benefit from the advantages that permissioned blockchains offer, for instance, decentralized trust. CRDT-enabled blockchains ease the realization of these use cases on blockchains.

\fabriccrdt, as an extension to \fabric, supports all use cases that can be implemented on \fabric. However, based on the requirements we explained in Section~\ref{fabric_requirments}, \fabriccrdt, by taking advantage of CRDTs, offers two additional properties which are beneficial to the CRDT use cases. \fabriccrdtspace ensures that (1) all submitted transactions that pass endorsement policy validation are committed successfully (\emph{no failure} requirement) and (2) no user updates are lost when concurrent updates on the same keys are submitted (\emph{no update loss} requirement), i.e., it offers eventual strong consistency.

One major use case that benefits from \fabriccrdtspace are collaborative document editing platforms, which provide an environment for users to concurrently work on shared documents. Because of the inherent concurrent nature of these platforms, conflicts from updating the same content can frequently occur. CRDTs are a practical technique for resolving these kind of conflicts~\cite{crdt_text_1, crdt_text_2}. By using CRDT features that \fabriccrdtspace offers, like JSON CRDTs, developers can create blockchain-based document editing applications. On \fabriccrdt, documents are stored as JSON objects, and edit updates are committed as CRDT transactions. Now, updates are merged without the loss of user's data (\emph{no update loss} requirement); further, no updates will fail, so that users do not need to redo and resubmit their edits (\emph{no failure} requirement). Furthermore, users benefit from the trust and security of permissioned blockchains when they use \fabriccrdt. Ref.~\cite{jsoncrdt} discusses how JSON CRDTs are used for representing text documents.

Another prominent application of permissioned blockchains is supply-chain management applications for tracing and ensuring the quality of different products from food to pharma industries~\cite{food_1, food_2}. Sensitive goods like drugs and fresh fruits and vegetables should be kept within specific conditions, e.g., regarding temperature, humidity, and light, during transportation and storage. To ensure that these goods are treated in compliance with regulations and policies, sensors continuously monitor the goods and record the readings on the blockchain to keep them secured against manipulations. Although the use case of storing a stream of sensor readings from IoT devices can be implemented on \fabric, we argue that this use case is even a better fit for \fabriccrdt. Depending on the design of the system, different readings from different IoT devices may collide, for example, when a temperature sensor and a humidity sensor concurrently submit records to update a shared list of the sensor readings of the same good. Using \fabriccrdt, it is ensured that conflicts are merged automatically and that all sensor data end up in the world state (\emph{no update loss} requirement). Due to resource limitations of IoT devices (e.g., regarding energy), the extra effort required for resubmitting failed transactions may be prohibitive. Using \fabriccrdtspace makes it possible for IoT devices to submit transactions \emph{once} without needing to take care of transaction failures and data loss (\emph{no failure} requirement).

There are also limitations to \fabriccrdt. Use cases that require transactional isolation of repeatable reads~\cite{transactions} are not a good fit, as \fabriccrdtspace commits transactions even if their read-set is outdated. This includes use cases for transferring assets. For example, financial applications like SmallBank~\cite{databasify} or FabCoin~\cite{mainfabric}, which are developed for \fabric, are bad choices to be adapted as a CRDT-based blockchain application. These applications represent asset creation and transfers between the owner; modeling them as CRDTs results in vulnerabilities, e.g., to the double-spending attack~\cite{double_spending}, where an attacker creates several transactions to transfer a single asset to multiple owners. On \fabric, only one of the attacker's transactions is successfully committed, and the MVCC validation fails on other transactions, since the committed transaction causes the read-set of other transactions to be outdated. However, \fabriccrdtspace skips the MVCC validation, merges the transactions' values, and successfully commits all of the attacker's transactions.

\section{Evaluation} \label{evaluation}

In this section, we provide a comprehensive evaluation of our design. We evaluate the number of successful transactions, latency, and the throughput of \fabriccrdtspace and \fabricspace under various configurations and workloads.

\subsection{Workload Generation} \label{workload}

Currently, the blockchain research community lacks a standard workload and benchmarking approach for evaluating different blockchain systems. Benchmarks such as TPC-C~\cite{TPC_C} and TPC-H~\cite{TPC_H} from the database community are not directly applicable to blockchains. They have been created for database systems and are not directly compatible with \fabricspace or \fabriccrdt. Adapting these workloads to the transactional structures of \fabricspace or other blockchain systems requires a steady community effort. 

We created a custom workload for evaluating the performance of \fabriccrdt, consisting of chaincodes for an IoT use case and use Hyperledger Caliper~\cite{caliper} for generating and submitting the transactions and collecting the performance metrics. When designing the workload, we focused on understanding the limitations and potentials of a CRDT-enabled \fabric. Since standard transactions in \fabricspace and \fabriccrdtspace go through identical workflows, we argue that for conflict-free workloads both systems show similar performance. For this reason, for most experiments, we evaluate the performance of \fabriccrdtspace on workloads that consist of conflicting read-write transactions. Additionally,  we perform one set of experiments with workloads consisting of conflicting and non-conflicting transactions in different ratios.

For our experiments, we implemented a chaincode that receives and stores temperature readings and device identification numbers of IoT devices. When executing a transaction, the chaincode first reads a key-value pair from the ledger, where the key is the device's identification number and the value is a JSON object containing the previous temperature readings of the device. Then, the chaincode adds the new temperature reading to the JSON object and submits it to be written to the ledger. As an example, Listing~\ref{lst:json_sample_submit} shows the JSON object that a transaction submits with one property for the device identification number and a list containing three temperature readings. For each experiment, the structure of the JSON object and the number of submitted JSON objects differ, which we will specify accordingly. However, the logic and behavior of the chaincode are the same for all experiments. 

\begin{lstlisting}[basicstyle=\small,label=lst:json_sample_submit,caption=Sample JSON object submitted by a transaction.]
{"deviceID": "e23df70a",
  "temperatureReadings": [
    { "temperature": 25 },
    { "temperature": 30 },
    { "temperature": 15 }
]}
\end{lstlisting}

\subsection{Experimental Setup} \label{setup}

We deployed a \fabricspace network on a Kubernetes v1.11.3 cluster consisting of three controller nodes, three worker nodes, one DNS and load balancer node, one NFS node, and one command-line interface (CLI) node. All nodes except the CLI node run on Ubuntu 16.04 virtual machines (VMs) with 16 vCPUs and 41 GB RAM. The CLI node runs on an Ubuntu 16.04 VM with 8 vCPUs and 20 GB RAM. All VMs are installed on top of KVM provided by OpenStack Mitaka and are interconnected by 10 GB Ethernet. We use CouchDB as the world state database and Apache Kafka/Zookeeper for the ordering service. Also, we use Hyperledger Caliper v0.1.0~\cite{caliper}. All experiments run a \fabricspace and a \fabriccrdtspace network of three organizations, two peers per organization, one orderer node, and one channel. 

For the evaluations, we kept the number of organizations, peers, channels, and clients constant. Since in \fabriccrdt, we did not change any \fabricspace components that are responsible for the communication between different parts over the network, we focus on evaluating the internal behavior of peers in \fabriccrdtspace and \fabric.

For each experiment, we start with an empty ledger and populate the ledger with keys that are read during the experiment as included in the configuration table of each experiment. During an experiment, Caliper uses four clients to submit in total 10,000 transactions. Besides this fixed setup, each experiment employs an additional configuration which we will specify.

\subsection{Effect of Different Block Sizes} \label{block_size}

We examine the impact of the block size on the total number of successful transactions, the throughput of successful transactions, and the average latency of successful transactions. We configured the \fabriccrdtspace and \fabricspace networks as described in Table~\ref{tab:blocksize_config} and only changed the block size in each experiment. We configured the maximum and preferred number of bytes for a block to 128 MB and the block timeout to 2 seconds. We kept these values fixed for each experiment but gradually increased the maximum allowed number of transactions in a block from 25 to 1000. Each chaincode invocation reads one key-value pair from the ledger and writes one pair back. Also, the JSON object that is written to the ledger has two keys, containing a string constant and a list, as we exemplify in Listing~\ref{lst:json_sample_submit}. 

In order to find the best configuration of \fabriccrdtspace and \fabricspace under worst-case workloads, all transactions modify the same keys; hence, all transactions are dependent on each other and are conflicting, \fabriccrdtspace will merge all the key-value pairs of all transactions in each block. Therefore, a higher number of transactions in a block induces a higher overhead for the peer to merge a higher number of JSON CRDTs.

\begin{table}[ht]
\caption{Configuration for evaluating the impact of the block size.}
\centering
  \begin{tabular}{ | p{6cm} | l |}
    \hline
    \textbf{Parameters}  & \textbf{Value}  \\ \hline
	Transaction submission rate per second  & 300 \\ \hline
	Number of read keys per transaction & 1 \\ \hline
	Number of write keys per transaction & 1 \\ \hline
	Number of keys per JSON object & 2 \\ \hline
  \end{tabular}
  \label{tab:blocksize_config}
\end{table}

\begin{figure}

  \begin{subfigure}{\linewidth}
  \centering
  \includegraphics[width=\linewidth]{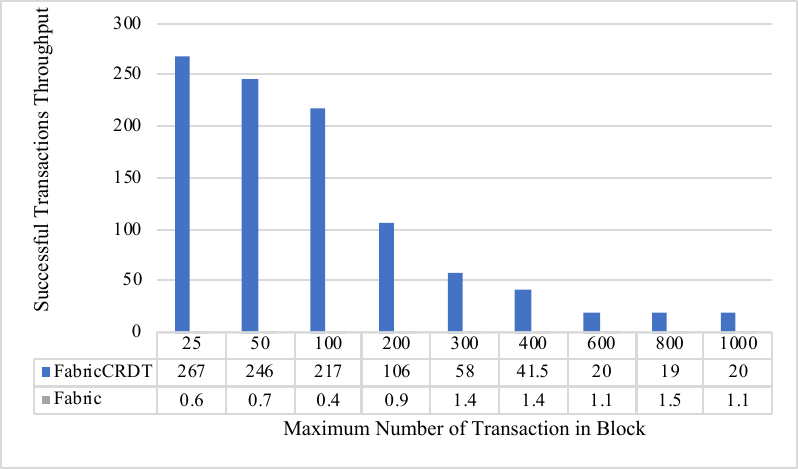}
  \caption{Successful transactions throughput per second for different block sizes.}
  \end{subfigure}\par\medskip
  
  \begin{subfigure}{\linewidth}
  \centering
  \includegraphics[width=\linewidth]{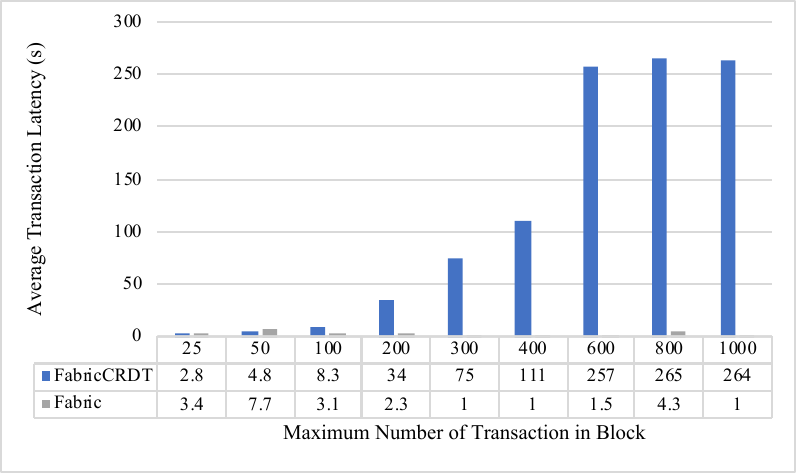}
  \caption{Average latency of successful transactions for different block sizes.}
  \end{subfigure}\par\medskip
  
  \begin{subfigure}{\linewidth}
  \centering
  \includegraphics[width=\linewidth]{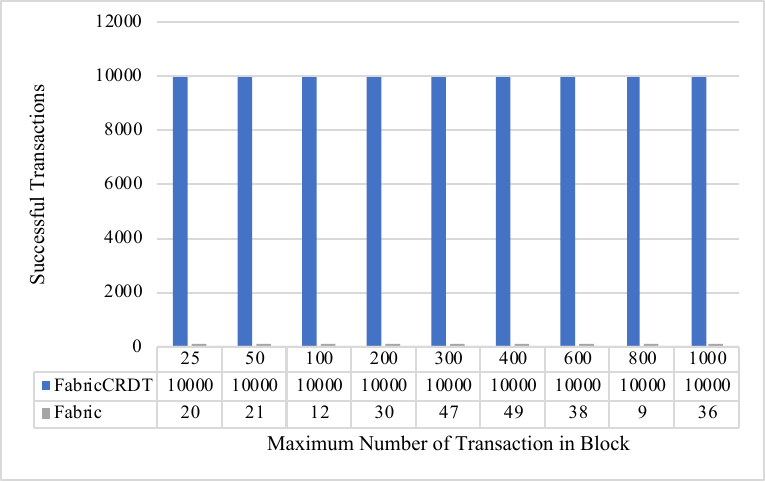}
  \caption{Number of successful transactions for different block sizes.}
  \end{subfigure}
  
  \caption{Effect of block size on throughput, latency, and success rate of \fabriccrdtspace and \fabric.}
  \label{fig:blocksize}
    
\end{figure}

\textbf{Observations -} The results of the experiments are summarized in Figure~\ref{fig:blocksize}. In Figure~\ref{fig:blocksize}(a), we can observe that \fabriccrdtspace has a higher throughput for smaller block sizes. The main reason for the degradation of throughput in \fabriccrdtspace with larger block sizes is the higher overhead required for merging a higher number of JSON CRDTs. For \fabriccrdt, the highest throughput overall was 267 transactions per second for a block size of 25 transactions. 

We can observe in Figure~\ref{fig:blocksize}(b) that \fabriccrdtspace experiences a higher transaction commit latency for larger block sizes because of the lower throughput, resulting in more time needed for committing all transactions. In Figure~\ref{fig:blocksize}(c), we can observe that \fabriccrdtspace successfully commits \emph{all} submitted transactions. In \fabric, there are always some conflicting transactions that cannot be committed, while in \fabriccrdt, all conflicts are automatically merged, and all transactions are successfully committed.

In the following experiments, we fix the block size to 25 transactions/block for \fabriccrdt, and to 400 transactions/block for \fabric. This way, we run both systems in their best configuration to get a fair comparison.

\subsection{Effect of Different Number of Reads and Writes} \label{var_keys}

To understand the effect of a higher number of key-value pairs in the transaction's read-write set, we change the number of key-value pairs that were read from and written to the ledger. For each experiment, we chose either 1, 3, or 5 key-value pairs to be read and to be written. Table~\ref{tab:keys_config} specifies the experimental configuration. During each experiment, we kept the set of read and write keys identical for all transactions. For example, in the experiment with five read-keys and five write-keys, in every transaction, we read or write the same set of 5 distinct keys.  

\begin{table}[ht]
\caption{Configuration for evaluating the impact of read and write keys.}
\centering
  \begin{tabular}{ | p{6cm} | l |}
    \hline
    \textbf{Parameters}  & \textbf{Value}  \\ \hline
	Transaction submission rate per second & 300 \\ \hline
	Number of keys per JSON object & 2 \\ \hline
  \end{tabular}
  \label{tab:keys_config}
\end{table}

\begin{figure}

  \begin{subfigure}{\linewidth}
  \centering
  \includegraphics[width=\linewidth]{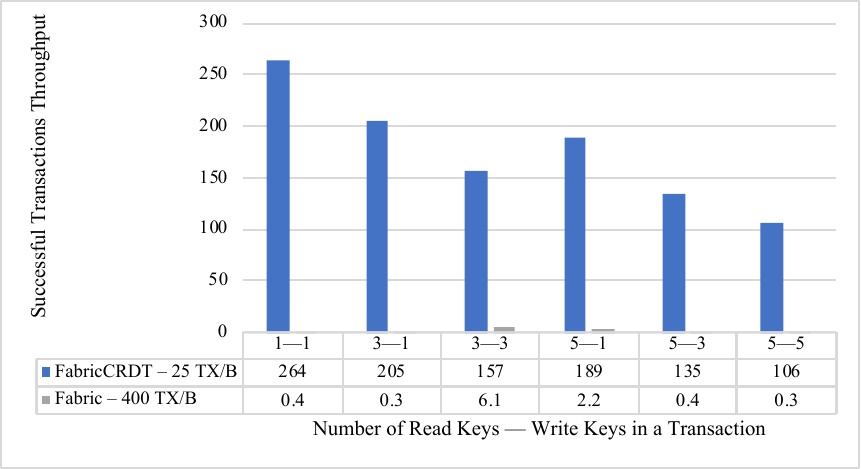}
  \caption{Successful transactions throughput per second for different number of read-write keys.}
  \end{subfigure}\par\medskip
  
  \begin{subfigure}{\linewidth}
  \centering
  \includegraphics[width=\linewidth]{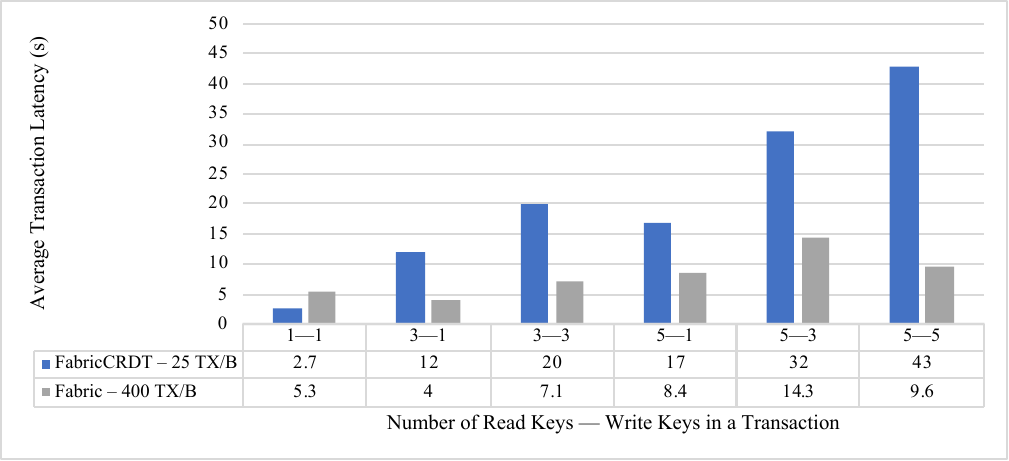}
  \caption{Average latency of successful transactions for different number of read-write keys.}
  \end{subfigure} \par\medskip
  
  \begin{subfigure}{\linewidth}
  \centering
  \includegraphics[width=\linewidth]{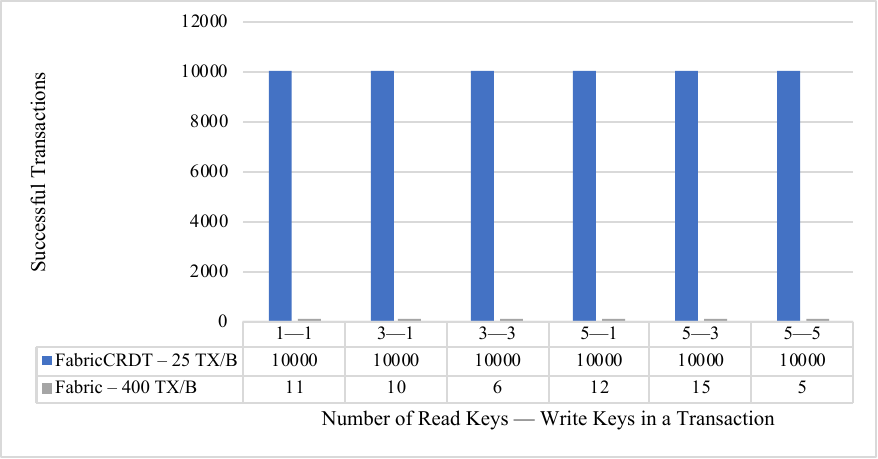}
  \caption{Number of successful transactions for different number of read-write keys.}
  \end{subfigure}
  
  \caption{Effect of different number of reads and writes in transactions on the throughput, latency, success rate of \fabriccrdtspace and \fabric.}
  \label{fig:key_values}
  
\end{figure}

\textbf{Observations -} Figure~\ref{fig:key_values} summarizes the results of the experiments. As expected, we can observe in Figure~\ref{fig:key_values}(a) that the throughput of \fabriccrdtspace decreases as the read-write set grows, because of the increased overhead for merging a larger number of values. We see that \fabriccrdtspace is affected by both the number of reads and writes in the transactions. In comparison to \fabriccrdt, \fabricspace shows a lower transaction throughput (Figure~\ref{fig:key_values}(a)) and a lower total number of successful transactions (Figure~\ref{fig:key_values}(c)). On the other hand, \fabriccrdtspace has a higher commit latency in comparison to \fabricspace (cf. Figure~\ref{fig:key_values}(b)).

\subsection{Impact of Varying Complexity of JSON Objects} \label{JSON_facror}

In contrast to the experiments in Section~\ref{var_keys}, here, we evaluate the effect of varying complexity of JSON objects that are written to the ledger. In particular, we study how the throughput and latency of \fabriccrdtspace changes, as merging more complex JSON objects induces more overhead. Table~\ref{tab:complex_config} shows the configuration of this experiment. Each transaction reads one JSON object from ledger with a certain number of keys and a certain nesting depth of the values; then, the transaction modifies the JSON object and writes it back to the ledger. Listing~\ref{lst:json_complex} exemplifies a JSON object with ``3-3 complexity'', i.e., the transaction submits a JSON object with three key-value pairs, where each value has a depth of three from the root of the JSON object.

\begin{table}[ht]
\caption{Configuration for evaluating the impact of different complexity of JSON objects.}
\centering
  \begin{tabular}{ | p{6cm} | l |}
    \hline
    \textbf{Parameters}  & \textbf{Value}  \\ \hline
	Transaction submission rate per second & 300 \\ \hline
	Number of read keys per transaction & 1 \\ \hline
	Number of write keys per transaction & 1 \\ \hline
  \end{tabular}
  \label{tab:complex_config}
\end{table}

\begin{figure}

  \begin{subfigure}{\linewidth}
  \centering
  \includegraphics[width=\linewidth]{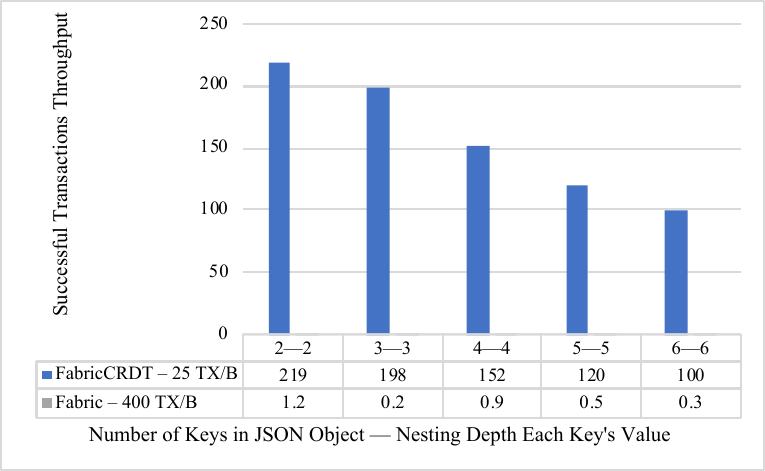}
  \caption{Successful transactions throughput per second for different JSON complexities.}
  \end{subfigure}\par\medskip
  
  \begin{subfigure}{\linewidth}
  \centering
  \includegraphics[width=\linewidth]{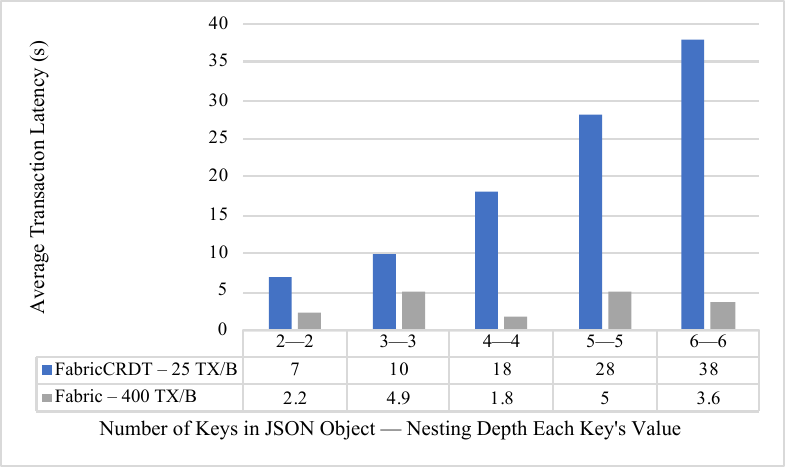}
  \caption{Average latency of successful transactions for different JSON complexities.}
  \end{subfigure}\par\medskip
  
  \begin{subfigure}{\linewidth}
  \centering
  \includegraphics[width=\linewidth]{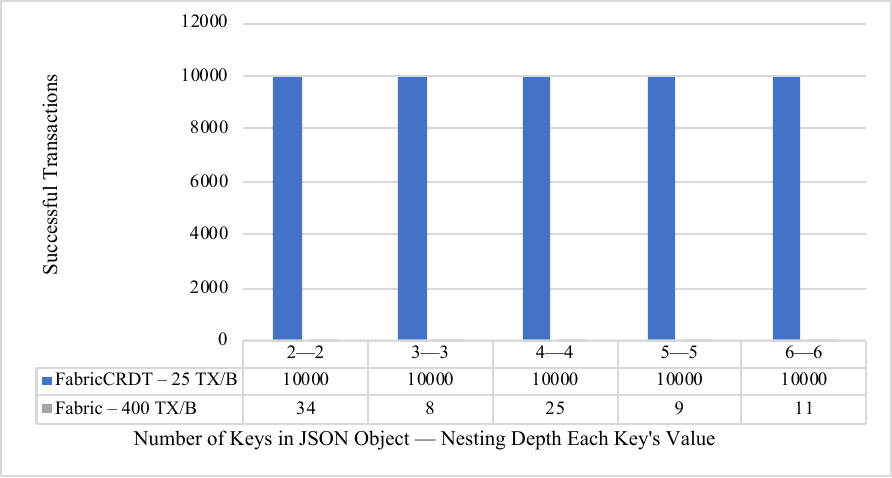}
  \caption{Number of successful transactions for different JSON complexities.}
  \end{subfigure}
  
  \caption{Effect of the complexity of JSON objects in the transactions on the throughput, latency and success rate of \fabriccrdtspace and \fabric.}
  \label{fig:json_values}
  
\end{figure}

\textbf{Observations -} Figure~\ref{fig:json_values} summarizes the results of the experiments. Similar to the experiments in Section~\ref{var_keys}, we observe that the throughput decreases and the latency increases for \fabriccrdtspace with an increasing complexity of JSON objects (cf. Figure~\ref{fig:json_values}(a) and Figure~\ref{fig:json_values}(b)). Unlike \fabriccrdt, \fabricspace does not interact with the content of the JSON objects. Therefore, the throughput and latency of \fabricspace are not correlated to the complexity of the JSON objects.

\begin{lstlisting}[basicstyle=\small,label=lst:json_complex,caption=A sample JSON object with ``3-3'' complexity.]
{"temperatureRoom1": [
    { "temperatureReading": [
        { "temperatureValue": 10
        }]}],
  "temperatureRoom2": [
    { "temperatureReading": [
        { "temperatureValue": 20
        }]}],
  "temperatureRoom3": [
    { "temperatureReading": [
        { "temperatureValue": 15
}]}]}
\end{lstlisting}

\subsection{Impact of Different Transaction Arrival Rates} \label{JSON_trans_arrival}

Further, we evaluate the effect of different transaction arrival rates on \fabriccrdtspace and \fabric. We configured each experiment according to Table~\ref{tab:tr_rate}. We employ four clients in total, where all clients together submit transactions with a rate of 100 to 500 transactions per second.

\begin{table}[ht]
\caption{Configuration for evaluating the impact of different transaction arrival rates.}
\centering
  \begin{tabular}{ | p{6cm} | l |}
    \hline
    \textbf{Parameters}  & \textbf{Value}  \\ \hline
	Number of read keys per transaction & 1 \\ \hline
	Number of write keys per transaction & 1 \\ \hline
	Number of keys in JSON objects & 2 \\ \hline
  \end{tabular}
  \label{tab:tr_rate}
\end{table}

\begin{figure}
  
  \begin{subfigure}{\linewidth}
  \centering
  \includegraphics[width=\linewidth]{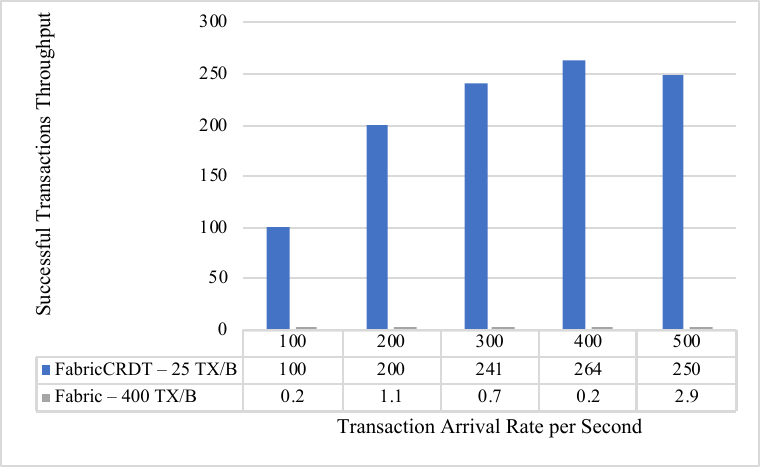}
  \caption{Successful transactions throughput per second for different transaction arrival rates.}
  \end{subfigure}\par\medskip
  
  \begin{subfigure}{\linewidth}
  \centering
  \includegraphics[width=\linewidth]{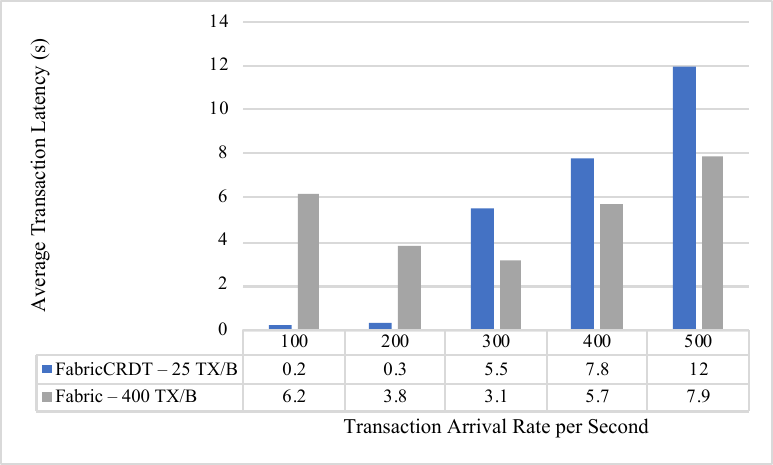}
  \caption{Average latency of successful transactions for different transaction arrival rates.}
  \end{subfigure}\par\medskip
  
  \begin{subfigure}{\linewidth}
  \centering
  \includegraphics[width=\linewidth]{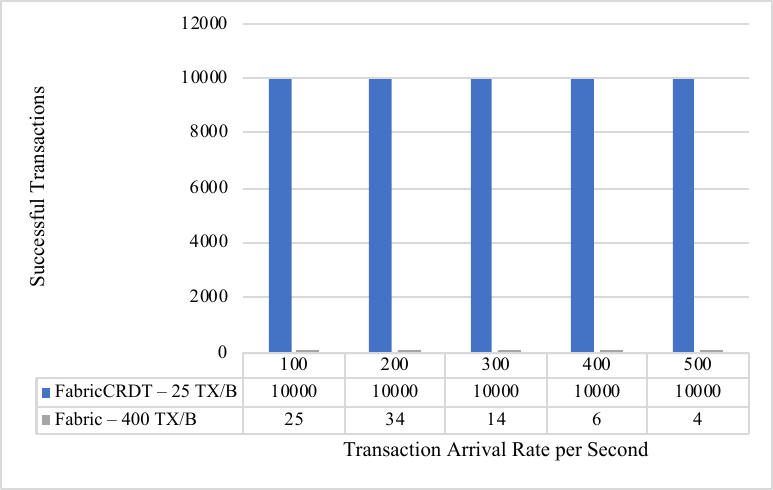}
  \caption{Number of successful transactions for different transaction arrival rates.}
  \end{subfigure}
  
   \caption{Effect of different transaction arrival rates on the throughput, latency, and success rate of \fabriccrdtspace and \fabric.}
  \label{fig:arrival_rate}
  
\end{figure}

\textbf{Observations -} As the results of the experiments in Figure~\ref{fig:arrival_rate}(a) show, \fabriccrdt's throughput increases until it reaches a saturation point at about 250 transactions per second. Meanwhile, Figure~\ref{fig:arrival_rate}(b) shows that the latency increases as the transaction arrival rate increases for \fabriccrdt. The enormous increase in latency in \fabriccrdtspace can be attributed to the effects of queuing when the transaction arrival rate exceeds the throughput.

\subsection{Impact of Different Percentage of Conflicting Transactions} \label{JSON_conflicting_rate}

In order to understand the limitations and potentials of \fabriccrdt, in the previous experiments, we used workloads where all transaction are conflicting. However, for production deployment of \fabricspace and \fabriccrdt, where different applications are hosted, blocks may contain both conflicting and non-conflicting transactions. To study the effects of different percentages of conflicting transaction in the workload, we configured experiments as specified in Table~\ref{tab:key_mixing}. For each experiment, a fixed percentage of transactions are conflicting, where the conflicting transactions are merged in \fabriccrdtspace and rejected in \fabric.

\begin{table}[ht]
\caption{Configuration for evaluating the impact of percentage of conflicting transactions in the workload.}
\centering
  \begin{tabular}{ | p{6cm} | l |}
    \hline
    \textbf{Parameters}  & \textbf{Value}  \\ \hline
	Transaction submission rate per second  & 300 \\ \hline
	Number of read keys per transaction & 1 \\ \hline
	Number of write keys per transaction & 1 \\ \hline
	Number of keys per JSON object & 2 \\ \hline
  \end{tabular}
  \label{tab:key_mixing}
\end{table}

\begin{figure}
  
  \begin{subfigure}{\linewidth}
  \centering
  \includegraphics[width=\linewidth]{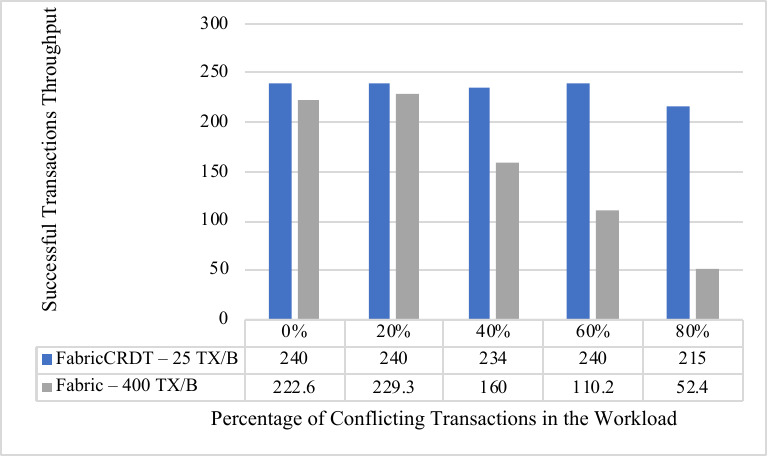}
  \caption{Successful transactions throughput per second for different percentage of conflicting transactions in the workload.}
  \end{subfigure}\par\medskip
  
  \begin{subfigure}{\linewidth}
  \centering
  \includegraphics[width=\linewidth]{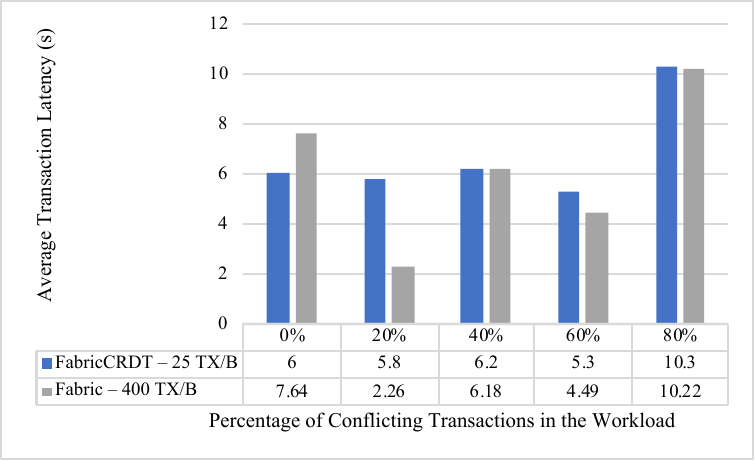}
  \caption{Average latency of successful transactions for different percentage of conflicting transactions in the workload.}
  \end{subfigure}\par\medskip
  
  \begin{subfigure}{\linewidth}
  \centering
  \includegraphics[width=\linewidth]{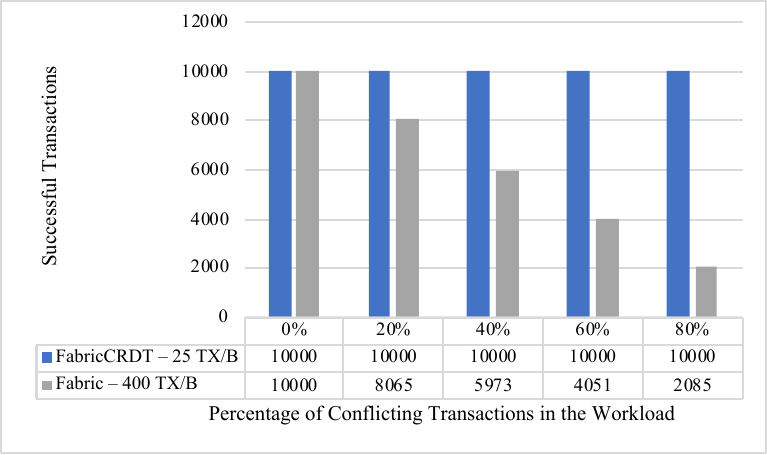}
  \caption{Number of successful transactions for different percentage of conflicting transactions in the workload.}
  \end{subfigure}
  
   \caption{Effect of different percentage of conflicting transactions in the workload on the throughput, latency, and success rate of \fabriccrdtspace and \fabric.}
  \label{fig:mixing_rate}
  
\end{figure}

\textbf{Observations -} Figure~\ref{fig:mixing_rate} summarizes the results of the experiment. We observe for workloads, where a smaller percentage of transactions are conflicting, that the throughput and latency of \fabriccrdtspace are similar to \fabricspace (Figure~\ref{fig:mixing_rate}(a) and Figure~\ref{fig:mixing_rate}(b)). However, when the percentage of conflicting transaction increases, the number of failures also increases in \fabricspace (Figure~\ref{fig:mixing_rate}(c)), while no failures occur in \fabriccrdt. 

\subsection{Summary of Results and Discussion} \label{discussion}

Since \fabriccrdtspace bypasses the MVCC validation and merges the conflicting transactions instead of rejecting them, it manages to commit all transactions in all experiments successfully. In stark contrast to this, \fabricspace only successfully a very few transactions when all transactions are conflicting. We argue that the performance of \fabriccrdtspace is an improvement to \fabric, since handling such a large amount of the transaction failures in the application may be a significant burden and increases the complexity of developing \fabricspace applications.
 
In our experiments, we observed that \fabriccrdt, in comparison to \fabric, suffers from a higher latency, which is a direct result of the extra processing required for merging a large number of JSON CRDTs. For adding each key in the JSON object to a JSON CRDTs, metadata has to be created, and the complexity of JSON CRDTs increases when more keys and values are added. However, in most parts of our evaluation, we investigated the worst-case scenarios where all transactions in a block are conflicting and are required to be merged. For scenarios where conflicting and non-conflicting transactions coexist, the results of experiments show that \fabricspace and \fabriccrdtspace have comparable latency and throughput.

\section{Related Work} \label{related_work}

Various CRDTs have been proposed and implemented in production-grade collaborative editing tools and distributed databases~\cite{roshi, crdt_seq, redis, Riak}. Although these works offer practical solutions for improving the scalability and enhancing the user experience, only a few works investigate the applicability of CRDTs on blockchains. In Ref.~\cite{dag_crdt}, the authors propose Vegvisir, a directed acyclic graph (DAG)-structured blockchain for CRDT-enabled applications. Vegvisir offers a power-efficient blockchain for IoT devices which tolerates network partitioning, but it only supports applications that can be implemented completely in CRDTs. Furthermore, a proposal has been introduced by \fabricspace developers to enhance \fabric's concurrency control by using built-in plugins for parallel execution of basic updates such as incrementing or decrementing values~\cite{enhanced_jira}. However, the implementation of this proposal has not been released, and the available information on the proposal is limited and lacks technical details.

Some works investigated various approaches for improving the performance and throughput of \fabric. In Ref.~\cite{databasify}, the authors use transaction reordering techniques~\cite{reordering} inspired by databases to improve the throughput of \fabricspace and to early abort conflicting transactions. They decrease the number of conflicting transactions by improving the order of the transactions in the ordering service according to a dependency graph. Although they show that reordering is a practical approach for decreasing transaction failures, they do not aim for the total elimination of failures, as \fabriccrdtspace does. 
Several works focused on identifying different bottlenecks of \fabricspace and offering solutions~\cite{streamchain, fastFabric, improved_ordering, FabricChar, performance}. In Ref.~\cite{streamchain, improved_ordering}, the authors offer solutions for improving the performance issues of the ordering service. The authors of StreamChain~\cite{streamchain} propose an approach for replacing \fabric's block processing mechanism with stream transaction processing to decrease the end-to-end latency of committing transactions. In Ref.~\cite{improved_ordering}, the authors propose a new Byzantine fault-tolerant protocol for the ordering service to increase the throughput of \fabricspace by decreasing the message communication overhead. The authors of Ref.~\cite{performance, fastFabric} offer extensive analysis and re-architecting guidelines of \fabricspace to improve several bottlenecks, including the consensus mechanism, I/O and computational overhead for ordering and validating transactions and repeated validation of certificates for endorsement policies. They implemented improvements such as the parallelization of several \fabricspace processes, separation of different resources, and caching. All these works provide valuable insights into different approaches that can improve the performance of \fabric. As \fabriccrdtspace reuses several of \fabric's components, these approaches are also applicable to \fabriccrdt. However, none of these works provide a solution for dealing with transaction failures of concurrent updates directly.

\section{Conclusion} \label{conclusion}

In this work, we introduced an approach for integrating CRDTs with
Hyperledger \fabric. We presented \fabriccrdt, an extension of \fabric,
that successfully commits transactions that perform concurrent updates
and automatically merges conflicting transactions by using CRDT
techniques without losing updates. We conducted extensive evaluations
to understand how \fabriccrdtspace performs in comparison to
\fabric. According to our findings, \fabriccrdtspace successfully merges all
conflicting transactions without any failures when all transactions
use CRDTs. In general, \fabriccrdtspace offers higher throughput than \fabricspace
but also induces higher commit latency due to the added overhead of
merging CRDTs.

In future work, we plan to extend \fabriccrdtspace with more CRDTs, such as
list, map, and graph CRDTs. We also investigate the effect of
eliminating the ordering service to understand the potentials of a
purely CRDT-based permissioned blockchain.

\begin{acks}
We would like to sincerely thank Kaiwen Zhang and our shepherd, Marko
Vukoli\'{c}, for their valuable input to this work. We also thank the
Alexander von Humboldt Foundation for supporting this project in part.
\end{acks}

\bibliographystyle{ACM-Reference-Format}
\bibliography{bibliography}

\end{document}